\documentclass[aps,prc,twocolumn,superscriptaddress,showpacs,floatfix,nofootinbib]{revtex4-1}

\usepackage{amsmath,graphicx,color}
\usepackage{graphicx} 
\usepackage{booktabs}
\usepackage{multirow}
\usepackage{graphics}
\usepackage{hyperref}
\usepackage{bm}
\usepackage{calligra}
\usepackage[T1]{fontenc}
\usepackage{amssymb}
\usepackage{amsfonts}
\usepackage{mathrsfs}
\usepackage{dsfont}
\usepackage{footmisc}
\usepackage[normalem]{ulem}
\usepackage{xcolor}
\usepackage{lineno}
\graphicspath{{figuresc/}}

\newcommand{\media}[1]{\left\langle #1 \right\rangle}
\newcommand\TRENTo{T\kern-.4ex\raisebox{-.48ex}{R}\kern-.40ex ENTo}

\begin{document}

\title{Universality of scaled particle spectra in ultrarelativistic heavy-ion collisions}

\author{Cicero D. Muncinelli}
\email{c250891@dac.unicamp.br}
\affiliation{%
  Universidade Estadual de Campinas (Unicamp), R. S\'ergio Buarque de Holanda, 777, Campinas, Brazil, 13083-859
}%

\author{Fernando G. Gardim}
\email{fernando.gardim@unifal-mg.edu.br}
\affiliation{Instituto de Ci\^encia e Tecnologia, Universidade Federal de Alfenas, 37715-400 Po\c cos de Caldas, MG, Brazil
}
\affiliation{%
  Illinois Center for Advanced Studies of the Universe \& Department of Physics, 
University of Illinois at Urbana-Champaign, Urbana, IL 61801-3003, USA
}%


\author{David D. Chinellato}
  \affiliation{Stefan Meyer Institute for Subatomic Physics
of the Austrian Academy of Sciences, Wiesingerstraße 4
1010 Vienna, Austria}

\author{Gabriel S. Denicol}
\affiliation{%
  Instituto de F\'isica, Universidade Federal Fluminense,
  Av. Milton Tavares de Souza, Niter\'oi, Brazil, 24210-346,
}%

\author{Andre V. Giannini}
\affiliation{%
  Faculdade de Ciências Exatas e Tecnologia, Universidade Federal da Grande Dourados, 
  Dourados, MS, Brazil, 79804-970
}
\affiliation{Departamento de F\'isica, Universidade do Estado de Santa Catarina, 89219-710 Joinville, SC, Brazil}


\author{Matthew Luzum}
\affiliation{
 Instituto de F\'isica, Universidade de S\~ao Paulo, R. do Mat\~ao, 1371, S\~ao Paulo, Brazil, 05508-090
}%

\author{Jorge Noronha}
\affiliation{%
  Illinois Center for Advanced Studies of the Universe \& Department of Physics, 
University of Illinois at Urbana-Champaign, Urbana, IL 61801-3003, USA
}%

\author{Tiago Nunes da Silva}
\affiliation{%
 Departamento de F\'{i}sica, Centro de Ciências Físicas e Matemáticas, Universidade Federal de Santa Catarina, Campus Universit\'{a}rio Reitor Jo\~{a}o David Ferreira Lima, Florian\'{o}polis, Brazil, 88040-900}%

\author{Jun Takahashi}
 \affiliation{%
  Universidade Estadual de Campinas (Unicamp), R. S\'ergio Buarque de Holanda, 777, Campinas, Brazil, 13083-859
}

\author{Giorgio Torrieri}
 \affiliation{%
  Universidade Estadual de Campinas (Unicamp), R. S\'ergio Buarque de Holanda, 777, Campinas, Brazil, 13083-859
}

\collaboration{The ExTrEMe Collaboration}

\date{\today}

\begin{abstract}
We study the transverse momentum spectra of identified particles in ultrarelativistic collisions of large and small collision systems. In order to isolate information contained in the momentum dependence, we propose to scale the spectra by the total particle number and mean transverse momentum --- global quantities which are already well studied.
We observe an interesting, nearly universal, centrality-independent shape in the scaled spectra, similar to scalings that have been studied previously.
This scaling behavior breaks down at large transverse momentum and for very small systems, such as those produced in p-p collisions.
We perform hybrid hydrodynamic simulations and show that, in these simulations, a centrality-independent shape is a consequence of an \emph{event-by-event} independence.
Our results motivate further theoretical and experimental investigations of the regime of validity of this scaling phenomenon and their physical interpretation at different collision energies and systems.
\end{abstract}


\maketitle

\section{Introduction}

High-energy nuclear collisions investigate the fundamental properties of nuclear matter under extreme conditions, shedding light on the nature of strong interactions. In these experiments, atomic nuclei collide at nearly the speed of light, producing an enormous number of particles; in the case of lead-lead collisions at the Large Hadron Collider (LHC), up to 35,000 particles can be produced. Studying the properties and correlations between these particles allows us to uncover key details about the behavior and dynamics of the matter created in these extreme conditions~\cite{Busza:2018rrf}. Experimental results from these collisions suggest that the matter produced behaves like a relativistic fluid made of quarks and gluons, flowing with very little kinematic viscosity~\cite{Heinz:2013th}, reaching temperatures billions of times hotter than the sun~\cite{Gardim:2019xjs}, in a volume less than $10^{-45}$ m$^3$, known as the quark-gluon plasma (QGP). Remarkably, in proton-lead collisions, where the reaction takes place in a volume one thousand times smaller than in lead-lead collisions, the matter produced still exhibits the characteristics of a QGP fluid when a large number of particles are produced~\cite{ALICE:2012eyl, CMS:2010ifv, Busza:2018rrf,Gardim:2022yds,Noronha:2024dtq}. 

One key observable for unraveling the collective behavior of the relativistic fluid produced at the collision's early stages is the distribution of particles in the plane transverse of the beam direction. In this regard, the azimuthal 
distribution of charged hadrons encoded in the flow coefficients $v_n$~\cite{Ollitrault:1992bk, Gardim:2012yp, Luzum:2013yya, ALICE:2016kpq}, has been the most insightful observable as it quantifies the final state momentum anisotropy of nuclear collisions. In fluid-dynamical models, this anisotropy emerges due to the hydrodynamic response to the initial-state geometry of the system, and its magnitude reflects how viscous the QGP is \cite{Luzum:2008cw}. 

Naturally, the azimuthally integrated transverse momentum distribution, $dN/dp_T$, must also carry relevant information about the fluid produced in these collisions. So far, this has been mostly explored using only two moments of this distribution: the total number of charged particles, $N_{ch}$, and the mean transverse momentum\footnote{To obtain $\media{p_T}$, additional modeling \cite{Schnedermann:1993ws,STAR:2008med,Guillen:2020nul} is used to extrapolate the spectrum down to $p_T=0$ GeV/c.}, $\media{p_T}$. These integrated observables are sensitive to the transport properties of the fluid, such as bulk viscosity \cite{Ryu:2015vwa}, and are usually included in global fits of hydrodynamic models through Bayesian analyses~\cite{Bernhard:2016tnd,Nijs:2020roc,JETSCAPE:2020mzn}. Nevertheless, it remains unclear what new information about the system's evolution is concealed in the full differential $p_T$-spectra.

This can be investigated by removing the global scales of total particle number, $N$, and mean transverse momentum, $\langle p_T\rangle$, isolating the shape of the spectra via the following scaled distribution:
\begin{equation}
 U\left(x_{T}\right)\equiv \frac{\media{p_T}}{N} \frac{dN}{dp_T}= \frac{1}{N} \frac{dN}{dx_T},
\label{eq:universal}
\end{equation}
\\
where $x_T=p_T/\media{p_T}$, following a similar idea to that proposed in Ref.~\cite{Hwa_2003:cent_pi}.

In this work we study this observable in a wide range of collision systems at LHC energies --- Pb-Pb collisions at $\sqrt{s_{NN}} = 2.76$ TeV and $\sqrt{s_{NN}} = 5.02$ TeV, p-Pb collisions at $\sqrt{s_{NN}} = 5.02$ TeV, Xe-Xe collisions at $\sqrt{s_{NN}} = 5.44$ TeV, and p–p collisions at $\sqrt{s} = 7$ TeV.
We perform a comprehensive set of hybrid hydrodynamic simulations of heavy-ion collisions and discover that, though the multiplicity and mean transverse momentum of each simulated event fluctuate significantly, the scaled spectrum remains almost identical.  This event-by-event universality survives in the event-averaged scaled spectra across centralities, leaving only a single universal curve.
This provides a distinct feature of fluid dynamical models that can be verified experimentally.

To do so, we gathered measured data from the ALICE Collaboration for each system and indeed observe a centrality independence, which begins to break down only at large transverse momentum and very small  systems (p-p).
These findings are consistent with various observed scalings that have been investigated over the past years \cite{Hwa_2003:cent_pi, Hwa_2003:high_pT, Hwa_2003:Mesons_pT_cent_energy, Yang:ppbar_AuAu, WANG201846:PbPb276, Zhang_2014:pp_ppbar_energy, Zhang_2015:pp_energy, Yang_2018:pp_strangeness, McLerran_2010:mult_meanpT, McLerran_2011:addendum, Zhu_2007:AuAu_dAu_cent, Zborovsk__2007:z_scaling, Hwa_2004:AuAu_cent_scaling, Liu_2019:pPb502_CSP, SONG2017516:pPb502_QCM, ZHU2008122:AuAu_cent, Fang_Lan_2008:dAu_rap, ZHU2008259:pp_mult_scaling, Tao_2022:Tsallis_scalings, Praszalowicz_2011:gsat_geom_scl, PRASZALOWICZ2013461:geom_scl, Praszalowicz_2011:geom_energy_cent, Mishra_2019:high_pT_pwr_law, Moriggi_2020:saturation_geom_scal, ALICE:light_flavor_mT_scaling, Khachatryan_2020:photon_scal, Gou_2017:QCM_pp_7TeV, Moriggi_2024:saturation_geom_scal_ALICE}.  Nevertheless, we remark that centrality-dependent scaling behavior for some of the systems studied here (Pb–Pb at 5.02 TeV, Xe–Xe at 5.44 TeV, and p–p at 7 TeV) has not been explored until now.

This paper is organized as follows. In Section \ref{sec:EbE}, we analyze the emergence of universal behavior in scaled spectra at an \emph{event-by-event} level in hybrid hydrodynamic simulations. In Section \ref{sec:Hydro_avg}, we extend this analysis to different collision systems and energies, further examining the scaling behavior as a function of centrality. Section \ref{sec:exp} presents a detailed study of experimental data from the LHC -- including Pb-Pb, Xe-Xe, p-Pb, and p-p collisions -- and investigates the extent to which scaling holds in different regimes. Finally, in Section \ref{sec:conclusion}, we summarize our findings and discuss their implications for understanding collectivity and the possible onset of hydrodynamic behavior across different systems.

\section{Event-by-event results in hybrid simulations}
\label{sec:EbE}

Fluid-dynamical models have been applied extensively to model and study the hot and dense matter produced in heavy-ion collisions. To understand the properties of the scaled transverse momentum spectra defined in Eq.~\eqref{eq:universal}, we first consider fluid-dynamical simulations for Pb-Pb collisions at 2.76 TeV, which involve: (i) initial conditions from the \TRENTo{} model~\cite{Moreland:2014oya}, (ii) a pre-equilibrium phase modeled using K\o MP\o ST \cite{Kurkela:2018wud}, (iii) 2+1 viscous hydrodynamics simulations performed using MUSIC~\cite{Schenke:2010nt, Paquet:2015lta} with the equation of state \cite{Huovinen:2009yb}, (iv) Cooper-Frye procedure \cite{Cooper:1974mv} and particlization via the hypersurface sampler iSS~\cite{Shen:2014vra}, and (v) hadronic transport modeling using UrQMD~\cite{Bass:1998ca}. The kinematic cuts employed are $p_T < 4.5$ GeV/c and $|y| < 0.5$ for rapidity. Detailed information about these simulations can be found in \cite{NunesdaSilva:2020bfs}. In the hybrid simulations, centrality is determined based on the total entropy of the initial conditions, whereas in experimental data, it is defined using the V0 scintillator hodoscopes located in the forward region ($2.8 < \eta < 5.1$ and $-3.7 < \eta < -1.7$)~\cite{pb276ALICE:2013mez, pb502ALICE:2019hno, ALICE:2021lsv, ppb502ALICE:2013wgn, pp7ALICE:2018pal}. The main advantage of first using simulations to address this scaling phenomenon is that it allows more detailed analyses, performed on an event-by-event basis, without relying on the transverse momentum spectra averaged over centrality classes. In the present work, 2334 minimum bias events were used, with oversampling. Each hypersurface was oversampled to a target number of 500,000 particles, and each event was averaged over the number of particlizations performed.

\begin{figure}[h]
    \centering

    \includegraphics[width=1.0\linewidth]{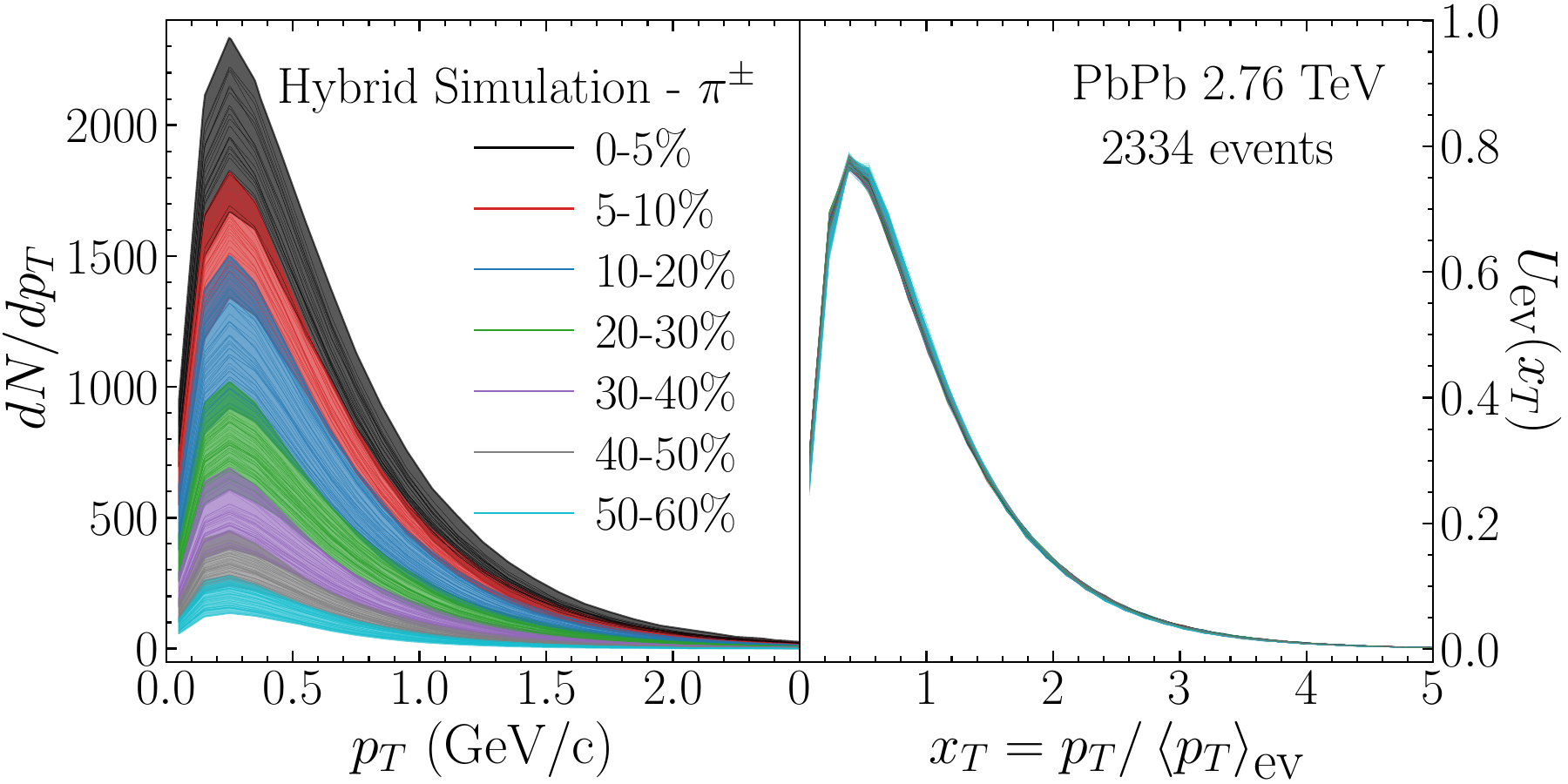} 
    \caption{(left) Presents the pion, $\pi^++\pi^-$, $p_T$-spectra for Pb-Pb at 2.76 TeV minimum bias, in the hybrid simulations. (right) Scaled spectra for each event, $U_{\rm ev}(x_T)=(\media{p_T}_{\rm ev}/N_{\rm ev})dN_{\rm ev}/dp_T$, computed using Eq.\ \ref{eq:universal}, for the events in the left plot. Centrality-independent universal behavior in $U_{\rm ev}(x_T)$ is observed.}
    \label{fig:motivation}
\end{figure}

Ultrarelativistic nucleus-nucleus collisions produce a variety of particles, with pions being the most abundant species comprising roughly 80$\%$ of the total yield~\cite{pb502ALICE:2019hno}. For the sake of simplicity, we present in this work only the results for charged pions, though we have checked that kaon and proton spectra display similar universal behavior, in agreement with \cite{WANG201846:PbPb276}. In Figure \ref{fig:motivation} (left), event-by-event $p_T$ spectra for Pb-Pb collisions at 2.76 TeV across different centralities are shown, distinguished by different colors. It is evident that $dN/dp_T$ changes with centrality and that there is a large variation within each centrality class. However, once the scaling proposed in Eq.\ \eqref{eq:universal} is applied to the $p_T$ spectra, all events approximately collapse onto the same curve, making them indistinguishable. We remark that the scaled spectra were constructed for \textit{each} event ($U_{\rm ev}$), as defined by Eq.\ \eqref{eq:universal}, incorporating the multiplicity of each event, $N_{\rm ev}$, and its corresponding average transverse momentum $\media{p_T}_{\rm ev}$. Thus, the scaled spectra shown in Fig.\ \ref{fig:motivation} (right) reveal that such universal behavior manifests itself for every single event, which could be a possible signature of a hydrodynamic phase. Additionally, we computed the standard deviation of $U_{\rm ev}$ for each $x_T$ and observed that the deviations are less than 5$\%$ up to $x_T = 4$.

The emergence of universal scaling in event-by-event hydrodynamic simulations is notable, particularly given the high complexity of modern hybrid models. Its presence implies that the spectral shape may be governed by a relatively simple mechanism even in these complex models. While similar scaling behaviors have been observed outside the hydrodynamic framework (see Ref.~\cite{WANG201846:PbPb276}), the fact that they arise here as well suggests that hydrodynamics can naturally reproduce this feature. If the underlying mechanism behind scaling in experimental data is indeed of hydrodynamic origin, the onset of such scaling may serve as a new experimental signature of fluid-like behavior in high-energy collisions.

\section{Varying systems and energy}
\label{sec:Hydro_avg}

\begin{figure}[h]
    \centering
    \includegraphics[width=\linewidth]{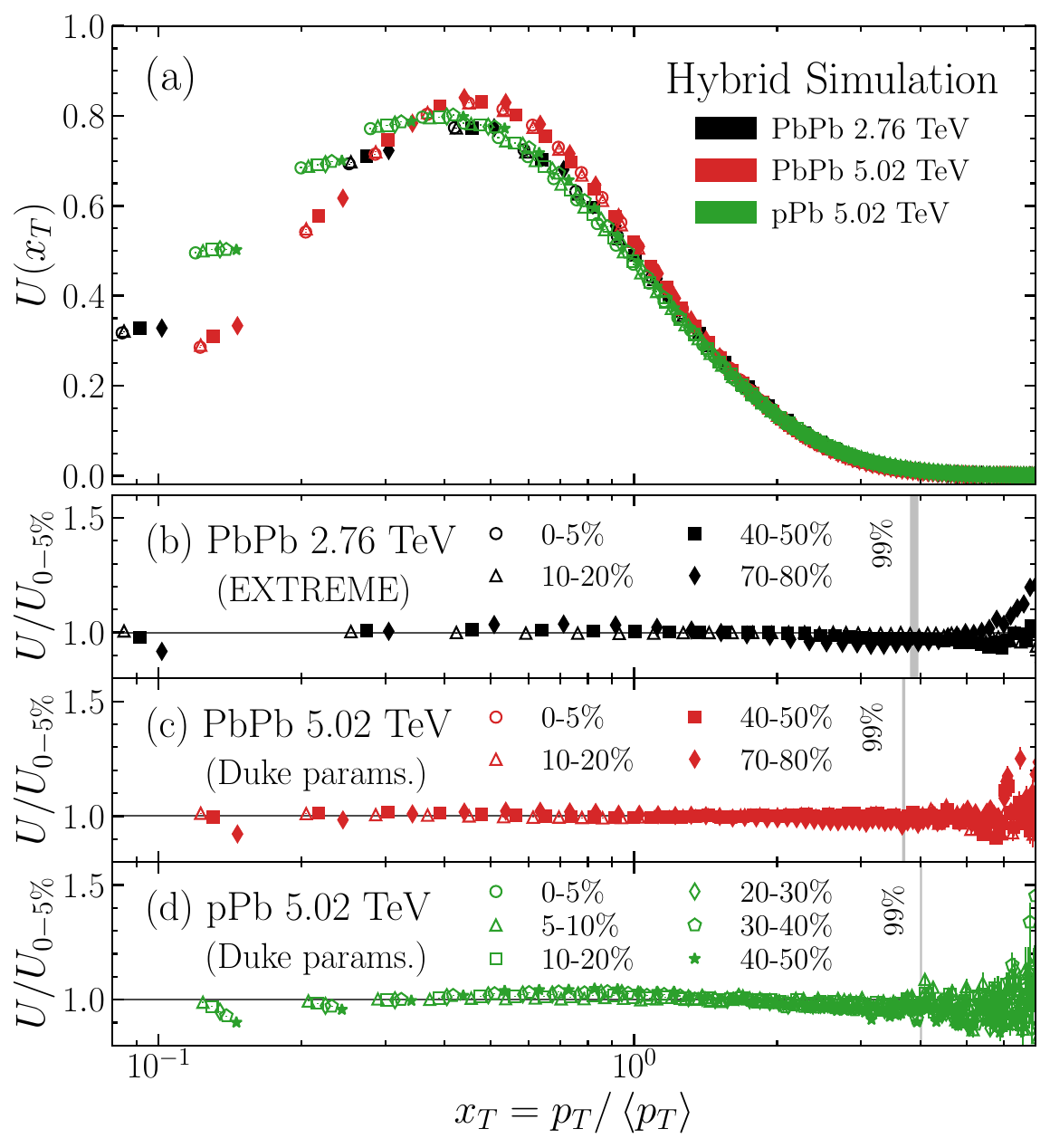} 
    \caption{(a) The scaled spectra, Eq.\ \eqref{eq:universal}, for $\pi^++\pi^-$ in Pb-Pb at 2.76 TeV and 5.02 TeV, and p-Pb 5.02 TeV, computed using hybrid simulations. Once the system is fixed, there is universality across centralities. (b) The ratio between $U$ coming from four centralities and $U$ for the most central collisions in Pb-Pb 2.76 TeV. (c) Ratio between $U$ coming from four centralities and $U$ for the most central collisions in Pb-Pb 5.02 TeV. (d) Ratio between $U$ coming from four centralities and $U$ for the most central collisions in p-Pb 5.02 TeV. We note that 99$\%$ of the pions are produced to the left of the vertical gray lines. The ExTrEME simulations (b) used the parameters from \cite{NunesdaSilva:2020bfs}, and the hybrid simulation presented in (c) and (d) used Duke parameters \cite{Moreland:2018gsh, daSilva:2022xwu}.}
    \label{fig:hydro}
\end{figure}

It is natural to wonder if the universality observed above holds for different energies and systems, as well as if it survives the averaging process over centrality classes employed in experiments. Furthermore, it is crucial to expand our analysis to include small systems, as there is an ongoing debate about whether the same features observed in large systems apply to small systems~\cite{Nagle:2018nvi,Noronha:2024dtq}. Given the consistent trend observed across events, we will focus on analyzing $U(x_T)$ per centrality, as this is the most accessible way to obtain experimental data. We verified this with two new sets of simulations of 1000 events each: p-Pb and Pb-Pb at 5.02 TeV, using parameters from~\cite{Moreland:2018gsh, daSilva:2022xwu} (see Figure \ref{fig:hydro}). It is important to note that in these simulations, the initial conditions are free-streamed into the hydrodynamic phase without using K\o MP\o ST, and centrality is defined based on the initial energy profile rather than the total entropy.
After fixing the nuclei and collision energy, we still observe agreement in $U$ across centralities, even for peripheral collisions of 70-80$\%$, as shown in the top panel of Figure \ref{fig:hydro}. Nevertheless, the shape of the universal distribution, $U$, varies when the model and collision system change. We use throughout this paper open symbols to represent centralities up to 40$\%$, and filled symbols for the other centralities, the peripheral ones. Results up to 90-100$\%$ centrality are not shown, but they do not affect our conclusions. 

To accurately compare the values of $U$ in different centralities, we computed the ratio between the $U$ distribution in each centrality class to that of the most central (0–5\%) bin. 
Different centrality bins have different $x_T$ binning (due to the different scaling factor $\langle p_T\rangle$), and so an interpolation between bins is necessary to make such a ratio.
Similar comparisons have been performed in Refs.~\cite{Yang:ppbar_AuAu, WANG201846:PbPb276, Zhang_2014:pp_ppbar_energy, Zhang_2015:pp_energy, Yang_2018:pp_strangeness, Zhu_2007:AuAu_dAu_cent, Liu_2019:pPb502_CSP, ZHU2008122:AuAu_cent, Fang_Lan_2008:dAu_rap, ZHU2008259:pp_mult_scaling, Tao_2022:Tsallis_scalings}, though typically by parametrizing the most central spectrum using a Tsallis function or other fitting forms. 
Here we instead choose a simple linear interpolation between $x_T$ bins for the 0-5\% centrality that is used as a denominator reference.

In the three bottom plots of Figure \ref{fig:hydro}, one can see that all $U$ curves are almost identical, with small deviations at very low and high $x_T$. These deviations are expected if the observed universality originates from hydrodynamic behavior, since the hybrid simulation incorporates decay and hadronic scattering, which can slightly modify some hydrodynamic signatures. For instance, at low $x_T$, particularly for pions, the spectra undergo some modifications due to decay. At high $x_T$, there is an additional contribution from hadron scattering~\cite{Schnedermann:1993ws}. The boundary covering 99$\%$ of pions produced is indicated by vertical lines, providing a reliable region where universality is observed in hydrodynamic models, regardless of energy and system size.

If systems of different sizes with equal collision energy exhibit the same multiplicity, it can be suggested that their $x_T$-spectra should be identical. This corresponds to the scenario involving Pb-Pb 70-80$\%$ and p-Pb 0-5$\%$ at 5.02 TeV. These collisions are represented by red diamonds and green circles in Figure \ref{fig:hydro} (a). It is evident that the corresponding $U(x_T)$ values for these collisions do not exhibit agreement. Having the same multiplicity does not necessarily mean having the same energy density and gradients, which means it will not lead to the same hydrodynamic expansion. 

Finally, we have examined the spectra of kaons and protons and reached similar conclusions. Further investigation of these particles will be presented in future work.

\section{Experimental results on the scaled spectra}
\label{sec:exp}

We now investigate the scaled transverse momentum spectra in the available experimental data, across various energies and systems.
In the case of heavy-ion collisions, we used ALICE Pb-Pb data at 2.76 TeV~\cite{pb276ALICE:2013mez} and 5.02 TeV~\cite{pb502ALICE:2019hno} as well as Xe-Xe data at 5.44 TeV~\cite{ALICE:2021lsv}. For p-Pb collisions at 5.02 TeV, we used ALICE data from~\cite{ppb502ALICE:2013wgn}, and for p-p collisions at 7 TeV we used ALICE data from~\cite{pp7ALICE:2018pal}. All experimental data used corresponds to the rapidity range of $|y|<0.5$, consistent with the range employed in the simulations. To obtain $U(x_T)$, it is essential to have information about the $p_T$-spectra down to 0 GeV/c, as the computation of the global quantities, $\media{p_T}$ and $N_{ch}$, relies on this information. Since there are no measurements for $p_T<0.1$ GeV/c for pions, further modeling is used to access these scales through extrapolations performed by the ALICE collaboration \cite{pb276ALICE:2013mez,ppb502ALICE:2013wgn}.

Before comparing the scaled spectra within the same system, it is important to emphasize that the experimental data contain physical effects that our simulations did not consider. This includes the impact of jets on the medium, which affects not only high-momentum particles, but also those with low momentum. Furthermore, the medium response to jets can differ due to variables such as size, centralities, and atomic mass number. Additionally, the $p_T$-spectrum for large momentum does not display thermal behavior and requires a description using a power-law function \cite{Hagedorn:1983wk, ALICE:2013txf}, indicating that it does not solely arise from hydrodynamics.

\begin{figure}[h!]
    \centering
    \includegraphics[width=\linewidth]{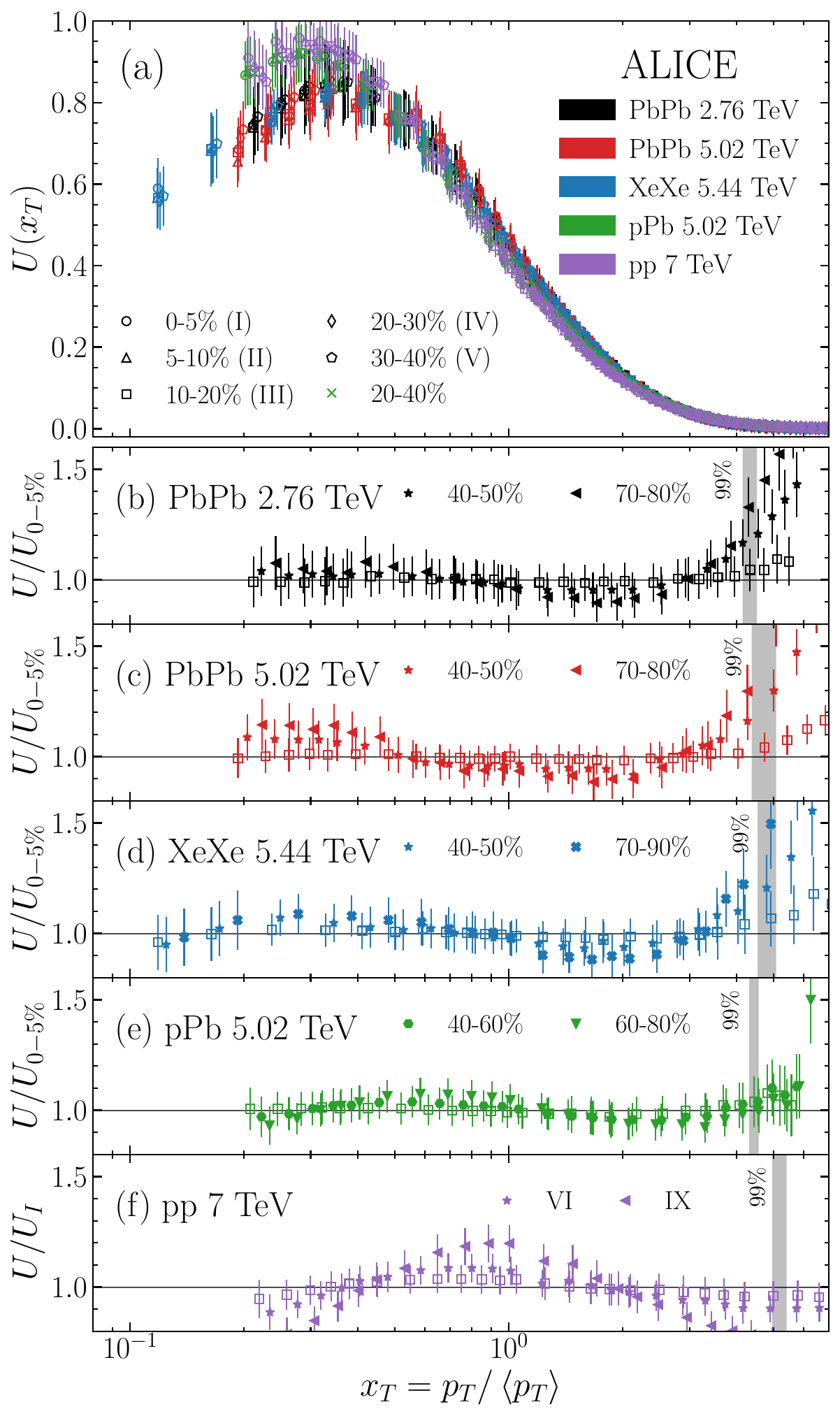}
    \caption{(a) The scaled spectra $U(x_T)$, as defined in Eq. \eqref{eq:universal}, for $\pi^+ + \pi^-$ in Pb-Pb at 2.76 TeV \cite{pb276ALICE:2013mez} and 5.02 TeV \cite{pb502ALICE:2019hno}, Xe-Xe 5.44 TeV \cite{ALICE:2021lsv}, p-Pb 5.02 TeV \cite{ppb502ALICE:2013wgn}, and p-p 7 TeV \cite{pp7ALICE:2018pal} from experimental data. The ratio between $U$ coming from four centralities and $U$ for the most central collisions is shown in (b) for Pb-Pb 2.76 TeV, (c) for Pb-Pb 5.02 TeV, (d) for Xe-Xe 5.44 TeV, (e) for p-Pb 5.02 TeV and (f) for p-p 7 TeV. 99$\%$ of the pions are produced to the left of the vertical gray lines. The p-p data is sorted by event activity, similar to centrality in heavy-ions, but with non-uniform percentile bin widths.  We follow ALICE and simply label them with Roman numerals I--IX.  See Table I of \cite{pp7ALICE:2018pal} for the corresponding percentiles.}
    \label{fig:alice}
\end{figure}

The results for different centralities, ranging from central to more peripheral regions, are presented in Figure \ref{fig:alice}, with error bars derived from the $p_T$-spectra and from the $\media{p_T}$'s and dN/dy's systematic errors from low $p_T$ extrapolations. For a fixed collision system and energy, represented by different colors in the plot, it is possible to observe the emergence of a universal $x_T$-spectrum, as shown in Figure \ref{fig:alice} (a). In the case of p-Pb, although universality is seen across centralities, $U(x_T)$ does not appear to be the same as in heavy-ion collisions, but we need smaller errors for $\media{p_T}$ to make a conclusive statement. The ALICE results, combined with our findings from hybrid simulations, suggest that universality may depend on energy and system size.

The comparison between $U(x_T)$ across centralities for the Pb-Pb and Xe-Xe measurements, as shown in Fig.~\ref{fig:alice} (b), (c), and (d), reveals universality (within experimental uncertainty). However, the agreement gradually diminishes as one goes to more peripheral collisions, indicating the potential onset of different physics. We further observe that the scaling behavior breaks down for large transverse momentum, where a hydrodynamic description is also expected to fail. This region may serve as an indication of where (semi-)hard physics begins to govern the production of particles.

The $U(x_T)$ spectra for small systems, p-Pb, and p-p, are shown in the plots (e) and (f) of Fig.~\ref{fig:alice}. For p-Pb, all centralities analyzed satisfy the aforementioned scaling behavior (within experimental uncertainty). On the other hand, in p-p collisions, even with large experimental error bars, it is evident that the scaling phenomenon breaks down as one deviates from high-multiplicity events. For instance, in centrality classes I, II, and III, which correspond to events within $\sim$ 0--10\%, the scaling behavior is manifest in p-p collisions. On the other hand, we already see significant deviations of the scaling for centrality class VI, which corresponds to $\sim$ 20--30\% \cite{pp7ALICE:2018pal}. Thus, the scaling phenomena investigated in this study suggest a potential connection to hydrodynamic behavior in p-Pb collisions, and possibly in high-multiplicity p-p collisions, offering insights into collectivity in small systems independently of anisotropic flow signatures.
However, given that the underlying mechanism remains unclear, further investigation is needed to substantiate this interpretation.

The behavior of $U(x_T)$ for kaons and protons in ALICE data was also examined, although not shown here, and the conclusions presented in this paper are unaffected.

\section{Conclusion}
\label{sec:conclusion}

In this work, we investigated a scaling behavior in the transverse momentum spectra of ultrarelativistic nuclear collisions, present both in experimental data and in event-by-event fluid-dynamical simulations. This was observed by rescaling the spectra, effectively removing the global scales of total particle number, $N$, and $\media{p_T}$, as defined in Eq.~\eqref{eq:universal}. We first demonstrated that this universal scaling appears in fluid-dynamical simulations of hadronic collisions, \emph{on an event-by-event basis}, even with the inclusion of nonhydrodynamic effects such as hadronic decays and rescattering. The scaling holds in the event-averaged spectra when the collision system and energy are fixed, although the shape of the reduced spectrum shows a slight evolution across systems. 

Using ALICE data, we computed the scaled spectra for charged pions in Pb-Pb collisions at 2.76 and 5.02 TeV, Xe-Xe collisions at 5.44 TeV, p-Pb collisions at 5.02 TeV, and in p-p collisions at 7 TeV. The same universal behavior  was observed in Pb-Pb and Xe-Xe collisions, for the bulk of produced particles. At high transverse momentum, the scaling breaks down, which could indicate a transition out of a hydrodynamics-dominated regime \cite{ALICE:2017nij, CMS:2021vui}. Combining these results with our hydrodynamic simulations, we hypothesize that the observed universality may serve as a signature of fluid-like behavior in high-energy collisions of large systems.

If this interpretation holds, it also offers new insights into current investigations of collectivity in small systems. In p-Pb and p-p collisions, where there is an ongoing debate about the formation of a QGP, we find using ALICE data that universal scaling holds for p-Pb collisions, but breaks down in p-p. This supports the possibility that a hydrodynamically-expanding QGP is formed in p-Pb events, and potentially even in high-multiplicity p-p events. In this regard, we note that our results are consistent with a previous study of the inferred thermodynamic properties of p-Pb collisions in the same regime \cite{Gardim:2022yds}. More definite conclusions would also require a reduction of the experimental uncertainty, in particular reducing systematic uncertainties in $\media{p_T}$ due to extrapolations. Extending measurements to lower transverse momenta would help in this regard.

Understanding the origin of the universality of scaled particle spectra in experimental data is crucial to determining whether it represents a robust signature of hydrodynamic behavior in QCD matter. The scaled spectrum can be further investigated in low center-of-mass-energy collisions to determine the importance of the hydrodynamic phase in such systems.
Furthermore, since hydrodynamic behavior is modified near a critical point  by the presence of additional slow degrees of freedom \cite{Stephanov:2017ghc}, it may be interesting to investigate the scaled spectrum also in the context of the ongoing search for the QCD critical point \cite{An:2021wof}.
Finally, our result can be instrumental to attest to the existence (or not) of hydrodynamic droplets in other systems, such as in $e^{+}-e^{-}$ collisions \cite{Badea:2019vey,Belle:2022fvl,Chen:2023njr}, or the presence of collective effects inside individual jets \cite{Baty:2021ugw,CMS:2023iam}. Such investigations may lead to new fundamental questions about the formation and evolution of hot and dense QCD matter in extreme conditions. 
Overall, our results motivate further theoretical and experimental investigations of this scaling phenomenon to bring to light the collective and non-collective behavior encoded in the transverse particle spectrum of different collision systems, expanding conventional particle physics studies, and improving our understanding of the QGP.

\begin{acknowledgments}

We thank J.-F.~Paquet for fruitful discussions. F.G.G.~is supported by CNPq (Conselho Nacional de Desenvolvimento Cientifico) through 306762/2021-8 and 307806/2025-1, and by the Fulbright Program. F.G.G.~and T. N.dS.~are supported by CNPq through the INCT-FNA grant 312932/2018-9 and the Universal Grant 409029/2021-1. M.L.~was supported by FAPESP projects 2017/05685-2, 2018/24720-6, and 2023/13749-1,  by project INCT-FNA Proc.~No.~464898/2014-5, and by CAPES - Finance Code 001.  G.S.D.~acknowledges support from CNPq and Funda\c c\~ao Carlos Chagas Filho de Amparo \`a Pesquisa do Estado do Rio de Janeiro (FAPERJ), grant No. E-26/202.747/2018. J.T.~was supported by FAPESP projects 2017/05685-2 and 2023/13749-1, and by CNPq through 303650/2025-7.  J.N. is partially supported by the U.S. Department of Energy, Office of Science, Office for Nuclear Physics under Award No. DE- SC0023861. C.D.M.~was supported by FAPESP projects 2017/05685-2, 2022/16166-4, 2023/13749-1, and 2025/01122-0, and by CNPq through the 122740/2023-8 and 119079/2024-0 grants.
G.T.~acknowledges support from Bolsa de produtividade CNPQ 305731/2023-8, Bolsa de pesquisa FAPESP 2023/06278-2.
\end{acknowledgments}

\bibliography{reff.bib}

\end{document}